# Adsorption-controlled growth of La-doped BaSnO$_3$ by molecular-beam epitaxy

Hanjong Paik,[1] Zhen Chen,[2] Edward Lochocki,[3] Ariel Seidner H.,[1] Amit Verma,[4] Nicholas Tanen,[1,5] Jisung Park,[1] Masaki Uchida,[6] ShunLi Shang,[7] Bi-Cheng Zhou,[7] Mario Brützam,[8] Reinhard Uecker,[8] Zi-Kui Liu,[7] Debdeep Jena,[1,5] Kyle M. Shen,[3,9] David A. Muller,[2,9] and Darrell G. Schlom[1,9,a]

[1] Department of Material Science and Engineering, Cornell University, Ithaca, New York 14853, USA

[2] School of Applied and Engineering Physics, Cornell University, Ithaca, New York 14853, USA

[3] Laboratory of Atomic and Solid State Physics, Department of Physics, Cornell University, Ithaca, New York 14853, USA

[4] Department of Electrical Engineering, IIT Kanpur, Kanpur 208016, India

[5] School of Electrical and Computer Engineering, Cornell University, Ithaca, New York 14853, USA

[6] Department of Applied Physics and Quantum-Phase Electronics Center (QPEC), University of Tokyo, Tokyo 113-8656, Japan

[7] Department of Materials Science and Engineering, The Pennsylvania State University, University Park, Pennsylvania 16802, USA

[8] Leibniz Institute for Crystal Growth, D-12489 Berlin, Germany

[9] Kavli Institute at Cornell for Nanoscale Science, Ithaca, New York 14853, USA

[a] Author to whom correspondence should be addressed. Electronic address: schlom@cornell.edu

**Abstract**

Epitaxial La-doped BaSnO$_3$ films were grown in an adsorption-controlled regime by molecular-beam epitaxy, where the excess volatile SnO$_x$ desorbs from the film surface. A film grown on a (001) DyScO$_3$ substrate exhibited a mobility of 183 cm$^2 \cdot$V$^{-1} \cdot$s$^{-1}$ at room temperature and 400 cm$^2 \cdot$V$^{-1} \cdot$s$^{-1}$ at 10 K, despite the high concentration (1.2×10$^{11}$ cm$^{-2}$) of threading dislocations present. In comparison to other reports, we observe a much lower concentration of (BaO)$_2$ Ruddlesden-Popper crystallographic shear faults. This suggests that in addition to threading dislocations that other defects—possibly (BaO)$_2$ crystallographic shear defects or point defects—significantly reduce the electron mobility.

Transparent conducting oxides with high mobility are being studied in hopes of realizing high-performance transparent electronics.[1] La-doped BaSnO$_3$ has emerged as a material of interest in this arena due to its high mobility at room temperature, transparency, and stability. La-doped BaSnO$_3$ single crystals are reported to have mobilities as high as 320 cm$^2 \cdot$V$^{-1}\cdot$s$^{-1}$ at room temperature at a mobile electron concentration of $n=8\times10^{19}$ cm$^{-3}$.[2] Indeed, La-doped BaSnO$_3$ has a higher mobility than all mainstream semiconductors (Si, GaAs, GaN, etc.) at doping concentrations above about $n=10^{19}$ cm$^{-3}$, where it is degenerately doped;[3] CdO is the only transparent semiconductor with higher mobility in this doping range.[4] Another advantage of BaSnO$_3$ is its excellent structural match to ferroelectric and antiferroelectric oxides with the perovskite structure, e.g., Pb(Zr,Ti)O$_3$. This could enable La-doped BaSnO$_3$ to serve as a high mobility channel for smart transistors[5] including ferroelectric field-effect transistors[6-16] and yield a subthreshold slope beating the 60 mV/decade Boltzmann limit of conventional field-effect transistors by fabricating negative capacitance field-effect transistors (NCFET).[17,18]

Two major deficiencies of today's epitaxially grown La-doped BaSnO$_3$ films that impact the performance of field-effect devices are (1) their mobility is significantly lower[2,3,19-24] than what has been demonstrated in La-doped BaSnO$_3$ single crystals[2,24,25] and (2) when doped below about $1\times10^{19}$ cm$^{-3}$ they are no longer conductive.[2,3,19-24] This latter point also applies to La-doped BaSnO$_3$ single crystals.[2,24] Both of these issues relate to the presence of significant concentrations of defects. The low mobility has been attributed to the high density of threading dislocations in epitaxial BaSnO$_3$ films that arise because they are grown on substrates to which they are poorly lattice matched.[2,19-24] High concentrations of threading dislocations are known to limit the mobility of other semiconductors including Ge,[26] (In,Ga)As,[27] In(As,Sb),[28] SiGe,[29] and GaN.[30] Indeed the mobilities of epitaxial GaN and BaSnO$_3$ films with threading dislocation densities in the $10^{10}$–$10^{11}$ cm$^{-2}$ range have been observed to scale with the square root of the mobile carrier concentration,[2,19,30] in agreement with theory.[26,30] In addition to the ability of dislocations to trap charge, nonstoichiometry, i.e., the ratio of (La+Ba):Sn deviating from 1 in La-doped BaSnO$_3$ films and the point defects it leads to, could also be responsible for the insulating behavior seen in lightly La-doped BaSnO$_3$ thin films. The inability to lightly dope La-doped BaSnO$_3$ layers is an obstacle to the fabrication of depletion-mode field-effect transistors.

The cutoff at about $1\times10^{19}$ cm$^{-3}$ in mobile electron concentration, below which doped films are insulating, is indicative of the concentration of electron traps in BaSnO$_3$ thin films. If nonstoichiometry is the root of the traps, then insulating behavior below a lanthanum concentration of $1\times10^{19}$ cm$^{-3}$ implies that the films deviate by 0.07% or more from being stoichiometric. This value is comparable to state-of-the-art stoichiometry control in the deposition of multicomponent films by physical vapor deposition methods.[31-39] A way to circumvent this limit is to exploit thermodynamics by entering an adsorption-controlled growth regime where the volatile constituents are provided in excess, but film composition is controlled automatically and locally through the volatility of those constituents to produce single-phase films.[40-49] Adsorption-control has been extensively used for the growth of oxides[50-52] including, most recently, for the growth of epitaxial BaSnO$_3$ films utilizing metalorganic precursors.[53]

In this Letter, we utilize adsorption-controlled growth with inorganic precursors to achieve La-doped BaSnO$_3$ thin films with (1) higher mobility and (2) that are conductive to lower carrier concentrations than have been reported to date. Room-temperature mobilities in excess of 150 cm$^2 \cdot$V$^{-1} \cdot$s$^{-1}$, the prior mobility record,[22] are achieved in fully relaxed La-doped BaSnO$_3$ thin films on substrates with mismatches ranging from −5.1% (SrTiO$_3$) to −2.3% (PrScO$_3$). Our result demonstrates that dislocations are not the only defect that limit the mobility in La-doped BaSnO$_3$ thin films and emphasizes the importance of precisely controlling film stoichiometry.

La-doped BaSnO$_3$ thin films were grown in a Veeco GEN10 MBE system from molecular beams emanating from separate effusion cells containing lanthanum (99.996% purity, Ames Lab), barium (99.99% purity, Sigma-Aldrich), and SnO$_2$ (99.996% purity, Alfa Aesar), respectively, in combination with a molecular beam of oxidant (the ~10% ozone + oxygen output of a commercial ozone generator).[54] The fluxes emanating from the effusion cells were determined by a quartz crystal microbalance (QCM) before growth. To achieve the desired doping concentration, the lanthanum flux was adjusted from the temperature at which its flux was measured by the QCM to a lower temperature, where accurate QCM measurements are not possible, by extrapolating its flux using the known activation energy of the vapor pressure of lanthanum,[55] i.e., a linear extrapolation of a plot of lanthanum vapor pressure vs. $1/T$. According to vapor pressure calculations, multiple species evaporate from SnO$_2$ under our growth conditions, with the major species being SnO.[56] In the supplementary material (S1) the calculated vapor pressure of species over solid SnO$_2$ are plotted at a fixed oxygen partial pressure of $7.6 \times 10^{-7}$ Torr ($10^{-9}$ atm). We used an excess of SnO$_x$-flux (above $9.0 \times 10^{13}$ atoms·cm$^{-2}$·s$^{-1}$) during growth, which is approximately three times greater than the barium flux ($3.0 \times 10^{13}$ atoms·cm$^{-2}$·s$^{-1}$). The background pressure of the oxidant, 10% O$_3$ + O$_2$, was held at a constant ion gauge pressure of $7.0 \times 10^{-7}$ Torr. All components—lanthanum, barium, SnO$_x$, and the 10% O$_3$ + O$_2$ oxidant—were co-supplied during film growth. A variety of perovskites substrates were used: (100) SrTiO$_3$, (001) DyScO$_3$, (110) DyScO$_3$, (110) TbScO$_3$, (110) GdScO$_3$, (110) Nd$_{0.5}$Sm$_{0.5}$ScO$_3$, (110) NdScO$_3$, and (110) PrScO$_3$.[57] These are all pseudocubic perovskite {100} surfaces and upon them the BaSnO$_3$ films grew with a cube-on-(pseudo)cube orientation relationship. The substrate temperature was maintained between 830-850 °C, as measured by an optical pyrometer. To determine the optimal single-phase growth window, we used *in situ* reflection high-energy electron diffraction (RHEED) as described below. The RHEED intensity oscillation period was used to estimate the film thickness and growth rate. The film growth rate was about 0.3 Å/s.

The phase purity and structural perfection of the films were assessed using four-circle x-ray diffraction (XRD) utilizing Cu $K_\alpha$ radiation with a high-resolution diffractometer (Panalytical X'Pert Pro MRD with a PreFix hybrid 4×Ge 220 monochromator on the incident beam side and a triple axis/rocking curve attachment (Ge 220) on the diffracted beam side). The microstructure and defects in the film were studied by cross-sectional and plan-view high (low)-angle annular dark field scanning transmission electron microscopy (HAADF-STEM and LAADF-STEM) with an FEI Titan Themis with a probe aberration corrector at 300 kV. Temperature-dependent electrical transport and Hall effect were measured in a van der Pauw geometry with contacts made by wire bonding.

Figure 1 shows the calculated oxygen partial pressure ($P_{O_2}$) vs. temperature ($T$) diagram for the Ba-Sn-O system with the tin partial pressure fixed at $7.6 \times 10^{-8}$ Torr ($10^{-10}$ atm). It is constructed using the CALPHAD method and first-principles calculations (see supplementary material for additional details (S2)).[58] The reaction enthalpy ($\Delta H$) values shown in Table I are used for the formation of $Ba_{n+1}Sn_nO_{3n+1}$ phases with $n=1, 2, 3$, and 4. The result is the four regions of stable solid phases shown in Fig. 1: (**I**) BaO, (**II**) $Ba_2SnO_4$, (**III**) $BaSnO_3$, and (**IV**) $SnO_2$, where the volatile $SnO_x$ gas phases are balanced with each solid phase. First-principles calculations indicate that there is no driving force to form $Ba_{n+1}Sn_nO_{3n+1}$ with $n > 2$;[59] hence the phases of $Ba_{n+1}Sn_nO_{3n+1}$ with $n > 2$ are not shown in Fig. 1; they are all lumped into stability region **II**. Overlaid onto Fig. 1 are RHEED patterns of La-doped $BaSnO_3$ thin films grown on (001) $DyScO_3$ substrates at different growth conditions (oxidant pressure and temperature).

Within region **III** stoichiometric $BaSnO_3$ films grow free of any surface reconstruction, i.e., with a 1×1 RHEED pattern. This can be clearly seen in Fig. 2(a) from the sharp 1×1 LEED image of a La-doped $BaSnO_3$ film. In contrast, we observe a 2×1 RHEED pattern, with the 2× reconstruction along the [110] azimuth of $BaSnO_3$ when the film growth conditions become slightly Ba-rich and move toward the boundary between region **III** and region **II** by either (1) increasing the substrate temperature, (2) lowering the flux supplied from the $SnO_2$ source, or (3) lowering the ozone partial pressure. Exiting region **III** and moving into region **II** is manifested by a more diffuse RHEED pattern with spots corresponding to the growth of a disordered Ruddlesden-Popper phase,[60-62] loaded with syntactic intergrowths of $Ba_{n+1}Sn_nO_{3n+1}$ layers with varying $n$. The $\theta$-$2\theta$ XRD pattern of a sample film grown in region **II** exhibiting such intergrowth disorder is shown in the supplementary material (Fig. S2). This pattern can be indexed as $Ba_8Sn_7O_{22}$. A hallmark of intergrowth disorder is the presence of both even and odd XRD indices;[63,64] an ideally ordered Ruddlesden-Popper phase would contain only even XRD indices because of the presence of the glide plane perpendicular to the $c$-axis.

If, on the other hand, starting from region **III** the fluxes are made more Sn-rich or the substrate temperature is lowered, a transmission RHEED pattern indicative of rough, three-dimensional growth is observed along both the [110] and [100] azimuths of $BaSnO_3$. This is indicative of the accumulation of $SnO_2$ in the film as the growth moves into region **IV**. The rough $SnO_2$ phase gives rise to the spots in the resulting RHEED pattern; the streaks are from the perovskite $BaSnO_3$ phase. The resulting mixed-phase sample corresponds to $SnO_2$+$BaSnO_3$ as shown by the XRD and RHEED results in the supplementary material (Fig. S5). Alternatively, if one again starts in region **III** and increases the ozone pressure (leaving all other growth

parameters constant) a three-dimensional transmission RHEED pattern indicative of condensed

$SnO_2$ on the film surface is seen. All of these observed changes are fully consistent with the

expectations implied by Fig. 1. The ability to see them *in situ* by RHEED allows one to reliably

find the desired growth window (region **III**) for the adsorption-controlled growth of phase-pure

$BaSnO_3$ thin films. For additional details see the supplementary material (S3).

Figure 2(b) shows RHEED intensity oscillations during the initial growth of a $BaSnO_3$ film on a (001) $DyScO_3$ substrate. The corresponding RHEED patterns of the same $BaSnO_3$ film along the [110] and [100] azimuths of $BaSnO_3$ are shown in Figs. 2(c) and 2(d), respectively.

The RHEED intensity oscillation was monitored at the off-specular position (marked by the red

box) along the [110] azimuth of $BaSnO_3$ shown in Fig. 2(c). Initially the $BaSnO_3$ film grew in a

layer-by-layer growth mode, but due to the large lattice mismatch (–4.2%) between the (001)

$DyScO_3$ substrate ($a_{DyScO_3,\text{pseudocubic}} = \sqrt{\frac{ab}{2}} = 3.943$ Å)[65] and $BaSnO_3$ film ($a_{BaSnO_3} =$

4.116 Å),[66] the film quickly relaxed and the amplitude of the RHEED oscillations decreased.

Concomitant with this relaxation, the growth mode changed to step-flow after the growth of about 13-15 unit cells. The film growth rate was 0.3 Å/s (equivalently ~0.1 μm/hr), based on both the RHEED intensity oscillations and thickness fringes observed by XRD.

The same $BaSnO_3$ film characterized by RHEED in Figs. 2(b)-2(d)—a 60 nm thick La-

doped $BaSnO_3$ film with a mobile carrier concentration of $1.2 \times 10^{20}$ $cm^{-3}$ grown on a 330 nm

thick undoped $BaSnO_3$ buffer layer on a (001) $DyScO_3$ substrate—is characterized by XRD in

Fig. 3. The $\theta$-$2\theta$ scan is shown in Fig. 3(a). The total film thickness is calculated based on the

Kiessig fringes[67] around the 002 Bragg peak of the $BaSnO_3$, as shown in Fig. 3(b). The $\theta$-$2\theta$ scan exhibits solely the *00ℓ* reflections of $BaSnO_3$ without any impurity phase. From these reflections the $c$-axis of this La-doped BaSnO$_3$ film is calculated to be $c$ = 4.116 ± 0.001 Å using a Nelson-Riley fit;[68] this is in agreement with the bulk lattice constant of BaSnO$_3$, $a$ = 4.116 Å.[66] A comparison of the structural perfection of this same La-doped BaSnO$_3$ film and the underlying DyScO$_3$ substrate it was grown upon are shown in Fig. 3(c). Here, the rocking curve of the 002 peak of the La-doped BaSnO$_3$ film is overlaid upon the 004 peak of the DyScO$_3$ substrate. The full width at half maximum (FWHM) of the film peak is 0.016°, which is far broader than the 0.0062° FWHM of the substrate. Although narrower than all prior reported FWHM for as-grown BaSnO$_3$–based heterostructures,[3,19,20,23,24] this relatively broad rocking curve is consistent with structural relaxation by the introduction of dislocations during the growth of the thick and highly mismatched (–4.2%) La-doped BaSnO$_3$ film on (001) DyScO$_3$. A reciprocal space map of the 103 BaSnO$_3$ peak of this same film is shown in Fig. 3(d). The in-plane and out-of-plane lattice constants of this La-doped BaSnO$_3$ film were calculated to be 4.1161±0.001 Å and 4.1163±0.001 Å, respectively, indicating that the La-doped BaSnO$_3$ film is fully relaxed. An atomic force microscope image of this same film is shown in the supplementary material (S4).

Figure 4 shows the temperature dependence of (a) resistivity, (b) carrier concentration, and (c) mobility of the same La-doped BaSnO$_3$ sample characterized in Figs. 2(b)-2(d) and 3. The resistivity at room temperature is 2.3×10$^{-4}$ Ω·cm and its temperature dependence exhibits metallic behavior down to 10 K with a resistivity ratio, $\rho_{300\,K} / \rho_{10\,K}$, of 2.15. The concentration of negatively charged mobile carriers ($n$) is temperature independent, as shown in Fig. 4(b). Assuming that all of the mobile carriers are attributable to the 60-nm-thick La-doped BaSnO$_3$ layer, the Hall resistance implies that $n$ is 1.2×10$^{20}$ cm$^{-3}$. The mobility ($\mu$) of this same sample was 183 cm$^2$·V$^{-1}$·s$^{-1}$ at room temperature and reached 400 cm$^2$·V$^{-1}$·s$^{-1}$ at 10 K as can be seen in Fig. 4(c). This room-temperature mobility is 20% higher than the previous record, 150 cm$^2$·V$^{-1}$·s$^{-1}$, which was achieved on a (110) PrScO$_3$ substrate.[22]

The sample described in detail so far, is our highest mobility sample. The room-temperature mobility of other La-doped BaSnO$_3$ samples grown using the same adsorption-controlled growth conditions on a variety of substrates and with differing doping concentrations are shown in Fig. 4(d). These substrates ranged from SrTiO$_3$ to PrScO$_3$, with lattice matches to BaSnO$_3$ ranging from –5.1% to –2.3%, respectively. Note that the room-temperature mobility of La-doped BaSnO$_3$ films on all of these substrates was higher than 160 cm$^2$·V$^{-1}$·s$^{-1}$ for doping concentrations in the (2–30)×10$^{19}$ cm$^{-3}$ range. Additionally, our growth conditions enable films with mobile carrier concentrations all the way down to 1×10$^{18}$ cm$^{-3}$ to be achieved;[69] this is an order of magnitude lower than prior reports.[2,3,19-24] The ability to dope BaSnO$_3$ at lower levels is consistent with the improved stoichiometry control that can accompany adsorption-controlled growth, leading to a reduction in the concentration of traps.

We investigated the defect structure of the La-doped BaSnO$_3$ sample with the highest mobility, the same sample whose other characteristics appear in Figs. 2-4, by STEM. A cross-sectional LAADF-STEM image of the entire film thickness is shown in Fig. 5(a). The high sensitivity of LAADF to strain and dislocations[70] makes it easy to see the threading dislocations. They are the vertically running defects with dark contrast in the BaSnO$_3$ film; one is indicated by a yellow arrow on Fig. 5(a). The HAADF-STEM images in Figs. 5(b) and 5(c) characterize the fully relaxed interface between the DyScO$_3$ substrate and the BaSnO$_3$ film. The spacing between the edge dislocations is on average 23 unit cells of DyScO$_3$ vs. 22 unit cells BaSnO$_3$, which is consistent with that calculated from the ratio of the relaxed lattice parameters. Extended dislocations can also be seen, as indicated by the yellow arrow in Fig. 5(b).

The density of threading dislocations in the same high-mobility sample characterized in Figs. 2-5 was determined by plan-view STEM measurements (Fig. 6) to be $1.2\times10^{11}$ cm$^{-2}$. A high-resolution HAADF-STEM image is shown in Fig. 6(d) showing two partial edge dislocations, each with Burgers vectors having in-plane projections of ½ *a* <110>. A full dislocation with a Burgers vector having an in-plane projection of *a*<110> is shown in the supplementary material (S5).

Interestingly, some of these dislocations have hollow cores. Being devoid of atoms, the hollow cores appear black in the plan-view HAADF-STEM images in Fig. 6(d) and Fig. S7 of the supplementary material. The magnitude of the smallest Burgers vector having an energetically stable hollow core lies in the range $20\pi\frac{\gamma}{\mu}$ to $40\pi\sqrt{e}\frac{\gamma}{\mu}$ for isotropic materials according to Frank's approximate theory,[71] where $\gamma$ is the surface energy and $\mu$ is the shear modulus. Using the calculated value of the surface energy (1.5 J/m$^2$)[72] and the measured value of the shear modulus (99.9 GPa)[73] of BaSnO$_3$, Frank's estimate of the minimum magnitude of the Burgers vector for it to have a hollow core lies in the 9-30 Å range. The two neighboring dislocations with outlined Burgers circuits in Fig. 6(d) both have Burgers vectors with in-plane projections of ½ *a* <110>, i.e., a magnitude $\frac{a}{\sqrt{2}}$ or 2.91 Å, yet one is hollow and the other is not. This could be because the out-of-plane component of the Burgers vectors of these two dislocations are not identical; they could have mixed character rather than being pure edge dislocations. Another possibility is that the adsorption controlled growth conditions lead to excess SnO$_x$ species on the film surface during growth, which acts as a flux that lowers $\gamma$.[71] The amount that $\gamma$ is lowered depends on the concentration of flux and could vary spatially, leading to dislocations that are hollow or not hollow even though they have identical magnitudes of their Burgers vectors.

The huge density of dislocations observed in this film with record mobility ($1.2 \times 10^{11}$ cm$^{-2}$) led us to question if there might be some other defects besides dislocations that currently limit mobility in BaSnO$_3$ films. After all, our films are grown on the same substrates and have comparable dislocation densities to prior studies,[19] yet the mobilities are far higher. How is it that our films have higher mobility? We do not know the answer to this question and are studying it further; what little we do know is mentioned below.

A potential culprit is Ruddlesden-Popper[60-62] (BaO)$_2$ crystallographic shear defects, which have been reported to be a dominant structural defect in La-doped BaSnO$_3$ films grown by pulsed-laser deposition.[74] The TEM images in the study of Wang *et al.*[74] reveal a concentration of (BaO)$_2$ crystallographic shear defects of about $2 \times 10^{11}$ cm$^{-2}$. In contrast, we see far fewer. We observed only one loop-shaped stacking fault in our highest mobility film (see the supplementary material (S6)). No stacking faults were observed in another two different areas with similar fields of view, leading us to estimate that the density of loop-shaped stacking faults in the film studied in Figs. 2-6 is about $3 \times 10^9$ cm$^{-2}$.

Differences in point defect concentrations could also be responsible for our films exhibiting higher mobility than other BaSnO$_3$ films with comparable dislocation densities. Vacancies on the barium site ($V_{Ba}''$) or on the tin site ($V_{Sn}''''$) are low-energy acceptor defects[75,76] in BaSnO$_3$ that could be responsible for the lack of conductivity in lightly La-doped BaSnO$_3$ films as well as the reduction in mobility when sufficient La is added to achieve conductivity. The local and automatic composition control provided by thermodynamics under adsorption-controlled growth conditions, could significantly reduce the concentration of $V_{Ba}''$, $V_{Sn}''''$, and other point defects, thus enhancing mobility. Note that adsorption-control is not synonymous with perfect composition control. Adsorption-control accesses the single-phase region of BaSnO$_3$, but depending on how wide that region is and from which side it is approached (in our case the SnO$_x$-rich side)—things that change with temperature and chemical potentials—the stoichiometry of the resulting film will change though it will always remain single phase. This is fully analogous to the growth of III-V compounds, where this behavior is well understood and utilized to controllably alter point defect concentrations, e.g., the EL2 defect in GaAs.[77]

In summary, using adsorption-controlled MBE growth, La-doped BaSnO$_3$ thin films with room-temperature mobilities as high as 183 cm$^2 \cdot$V$^{-1} \cdot$s$^{-1}$ were achieved on highly mismatched substrates, despite high concentrations (~10$^{11}$ cm$^{-2}$) of threading dislocations. Further, this growth method enabled La-doped BaSnO$_3$ with mobile carrier concentrations as low as $1 \times 10^{18}$ cm$^{-3}$ to be achieved.[69] These results imply that threading dislocations are not the only defects that have been limiting the mobility and trapping carriers in La-doped BaSnO$_3$ thin films. Other defects, possibly (BaO)$_2$ crystallographic shear defects or point defects arising from non-stoichiometry, are potential culprits. These results make us believe that the combination of adsorption-controlled MBE with lattice-matched perovskite substrates will be a promising path to high-mobility La-doped BaSnO$_3$ thin films.

See supplementary material for additional details regarding the thermodynamic calculations as well as the structural and spectroscopic characterization of the BaSnO$_3$ films.

We gratefully acknowledge stimulating discussions with Karthik Krishnaswamy and Chris Van de Walle. This material is based upon work supported by the Air Force Office of Scientific Research under award number FA9550-16-1-0192 and by the National Science Foundation (Platform for the Accelerated Realization, Analysis, and Discovery of Interface Materials (PARADIM)) under Cooperative Agreement No. DMR-1539918. We also acknowledge support from the Center for Low Energy Systems Technology (LEAST), one of the six SRC STARnet

Centers, sponsored by MARCO and DARPA. This work made use of the Cornell Center for Materials Research (CCMR) Shared Facilities, which are supported through the NSF MRSEC program (DMR-1120296). Substrate preparation was performed in part at the Cornell NanoScale Facility, a member of the National Nanotechnology Coordinated Infrastructure (NNCI), which is supported by the NSF (Grant ECCS-15420819).

**Table I.** Reaction enthalpy ($\Delta H$) values for the formation of $Ba_{n+1}Sn_nO_{3n+1}$ phases with $n$=1, 2, 3, and 4, calculated from first-principles with the PBEsol functional.

| Reaction | $\Delta H$ (eV/atom) | $\Delta H$ (kJ/mol-f.u.) |
|---|---|---|
| $BaO + SnO_2 = BaSnO_3$ | −0.223 | −107.5 |
| $2BaO + SnO_2 = Ba_2SnO_4$ | −0.228 | −154.2 |
| $3BaO + 2SnO_2 = Ba_3Sn_2O_7$ | −0.227 | −262.4 |
| $4BaO + 3SnO_2 = Ba_4Sn_3O_{10}$ | −0.225 | −369.3 |

**Figure 1.** Calculated Ellingham diagram (oxygen partial pressure vs. reciprocal temperature with the tin partial pressure fixed at $7.6\times10^{-8}$ Torr ($10^{-10}$ atm) assuming an open system. The overlaid RHEED patterns are taken along the [110] $BaSnO_3$ azimuth from films grown on (001) $DyScO_3$ substrates at different substrate temperatures. The four regions of phase stability between $SnO_x$ gases and the condensed phases are represented as (**I**) BaO, (**II**) $Ba_2SnO_4$, (**III**) $BaSnO_3$, and (**IV**) $SnO_2$, respectively, where the name of each region corresponds to the major condensed phase present. First-principles calculations, using the generalized gradient approximation (GGA) with Perdew–Burke–Ernzerhof revised for solids (PBEsol) functional, predicted the enthalpy of $BaSnO_3$ formation to be -107.5 kJ/mol per formula unit for the $BaO+SnO_2=BaSnO_3$ reaction (see Table I).

**Figure 2.** (a) Low-energy electron diffraction (LEED) pattern of a 25 nm thick, 3.5 at. % La-doped $BaSnO_3$ thin film grown on a (110) $TbScO_3$ substrate. (b) Reflection high-energy electron diffraction (RHEED) intensity oscillation during the growth of an undoped $BaSnO_3$ buffer layer

on a (001) DyScO$_3$ substrate. RHEED images of a 60 nm thick La-doped BaSnO$_3$ film with a mobile carrier concentration of $1.2 \times 10^{20}$ cm$^{-3}$ (grown on top of the 330 nm thick undoped BaSnO$_3$ buffer layer shown in (b)) viewed along the (c) [110] and (d) [100] azimuth of BaSnO$_3$.

**Figure 3.** XRD scans of a 60 nm thick La-doped BaSnO$_3$ film grown on a 330 nm thick undoped BaSnO$_3$ buffered layer on a (001) DyScO$_3$ substrate measured in a triple-axis geometry. (a) $\theta$-$2\theta$ scan. (b) A close-up view of the $\theta$-$2\theta$ scan around the 002 La-doped BaSnO$_3$ peak showing clear thickness fringes. The total thickness of the BaSnO$_3$ film is calculated to be $390 \pm 0.2$ nm. (c) Overlaid rocking curves of the 002 BaSnO$_3$ film peak and the 004 DyScO$_3$ substrate peak. (d) A reciprocal space map around the 103 BaSnO$_3$ film and the 332 DyScO$_3$ substrate peak. The substrate peaks are labeled with asterisks.

**Figure 4.** (a) Resistivity vs. temperature, (b) mobile electron carrier concentration vs. temperature, and (c) electron mobility vs. temperature of the same La-doped BaSnO$_3$ film characterized in Figs. 2(b)-2(d) and 3. In (d) measurements of the mobility vs. mobile electron carrier concentration are made for a multitude of La-doped BaSnO$_3$ films grown on (100) SrTiO$_3$, (001) DyScO$_3$, (110) DyScO$_3$, (110) TbScO$_3$, (110) GdScO$_3$, (110) Nd$_{0.5}$Sm$_{0.5}$ScO$_3$, (110) NdScO$_3$, and (110) PrScO$_3$ substrates. All of the "Cornell" films were grown under the adsorption-controlled growth conditions described. Also plotted for comparison are the highest mobility La-doped BaSnO$_3$ single crystals from Kim et al.[2] at Seoul National University (SNU, solid blue squares) and the highest mobility La-doped BaSnO$_3$ films from Raghavan et al.[22] at the University of California, Santa Barbara (UCSB, green triangle),

Kim *et al.*[2,24] at SNU (purple diamond), Shiogai *et al.*[21] at Tohoku University (orange upside down triangle), and Prakash *et al.*[23] at the University of Minnesota (cyan sideways triangle).

**Figure 5.** Cross-sectional STEM images of the same La-doped $BaSnO_3$ film characterized in Figs. 2-4. (a) LAADF-STEM image showing the entire film thickness. The yellow arrow indicates a threading dislocation. HAADF-STEM images of the $BaSnO_3$/$DyScO_3$ interface are shown in (b) and (c). Edge dislocations are labeled on (c).

**Figure 6.** Plan-view STEM images of the same La-doped $BaSnO_3$ film characterized in Figs. 2-5. (a) Bright-field and (b) dark-field STEM images. (c) and (d) are low and high magnification HAADF-STEM images, respectively. From these images the density of threading dislocations is $1.2\times10^{11}$ cm$^{-2}$. The yellow arrow in (a) shows a threading dislocation. Four dislocations are present in (d). The Burgers circuit is drawn for the two on the left, revealing two partial edge dislocations, each with a Burgers vector with an in-plane projection of ½ *a* <110>. The dislocation that is arrowed is not hollow, whereas the dislocation below it is hollow.

Figure 1.

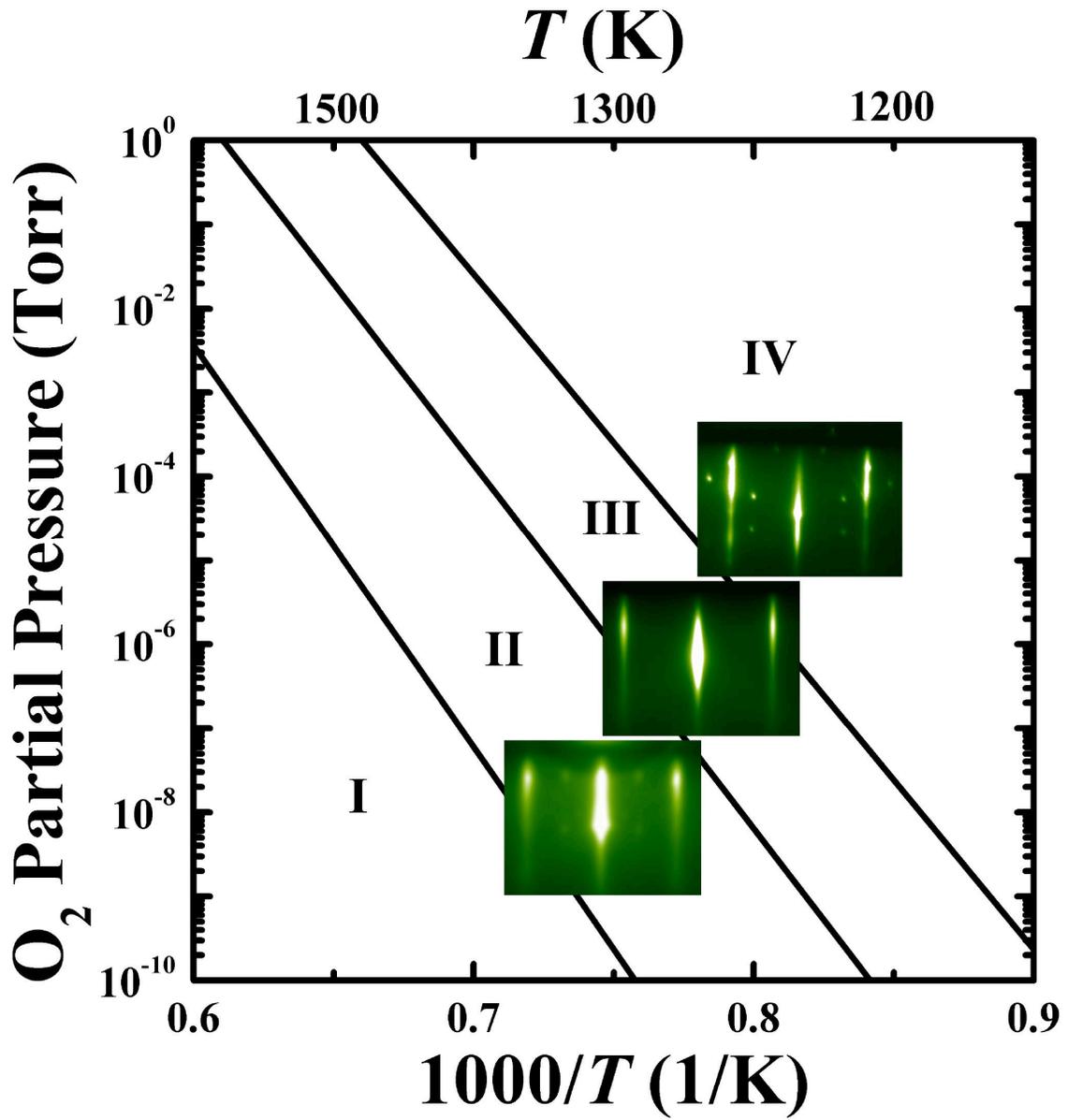

**Figure 2.**

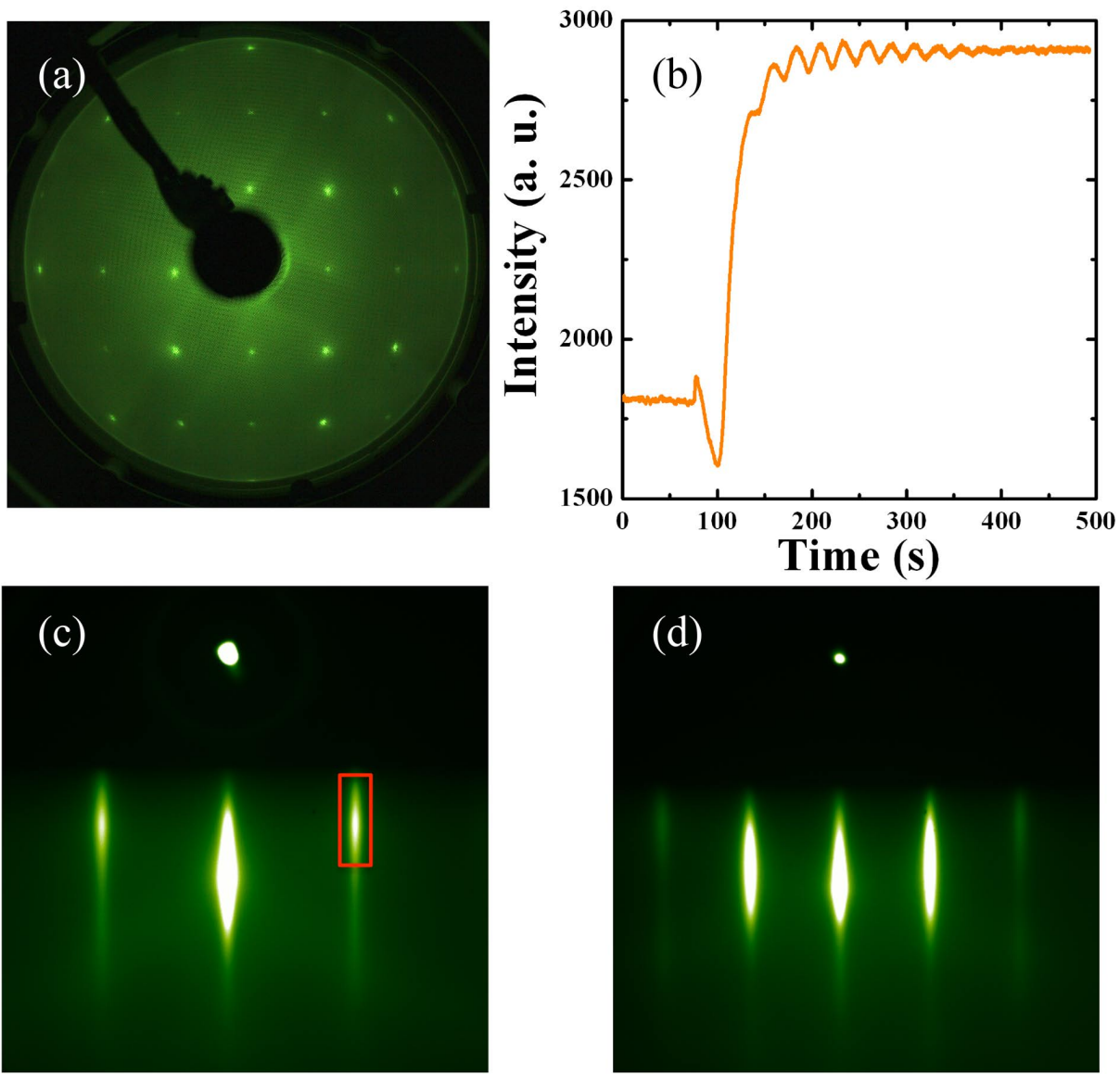

**Figure 3.**

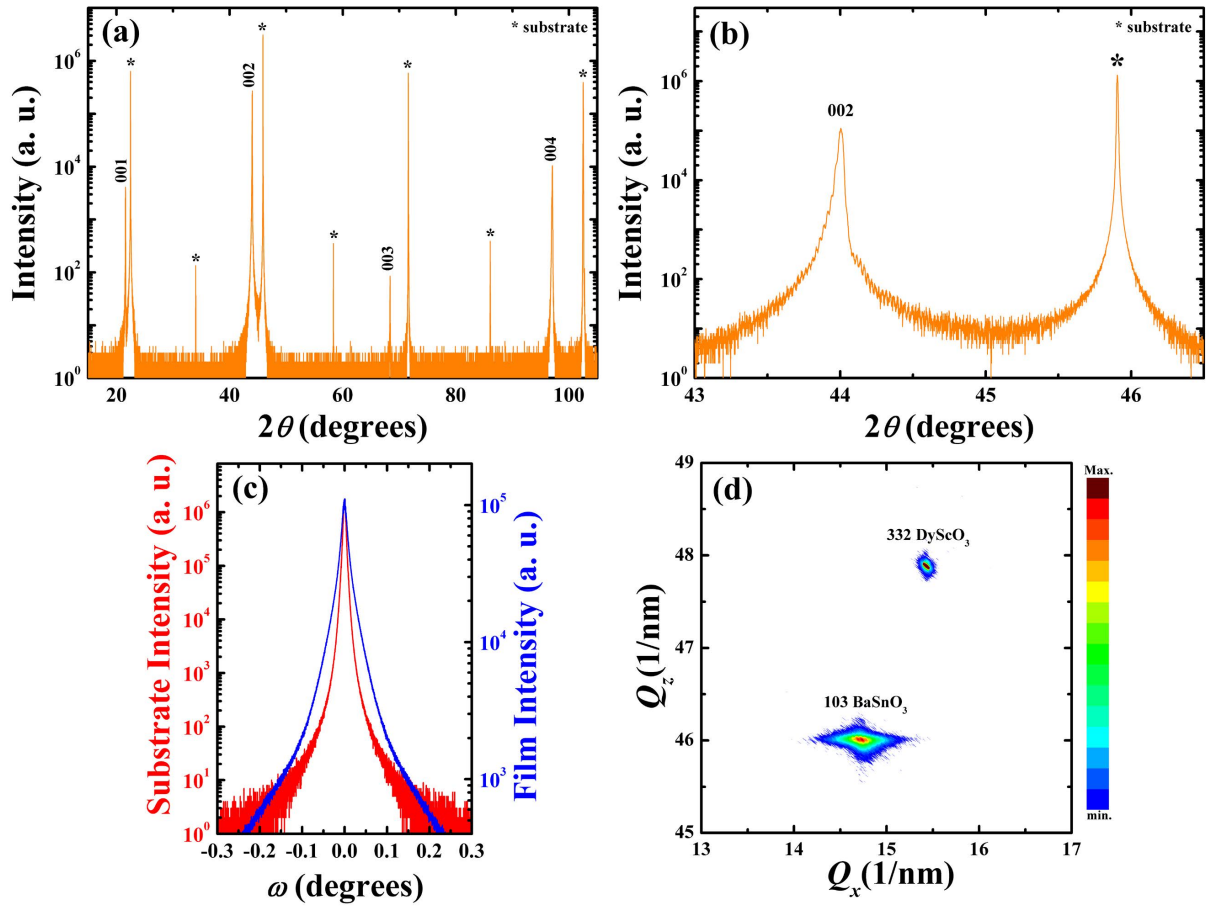

**Figure 4.**

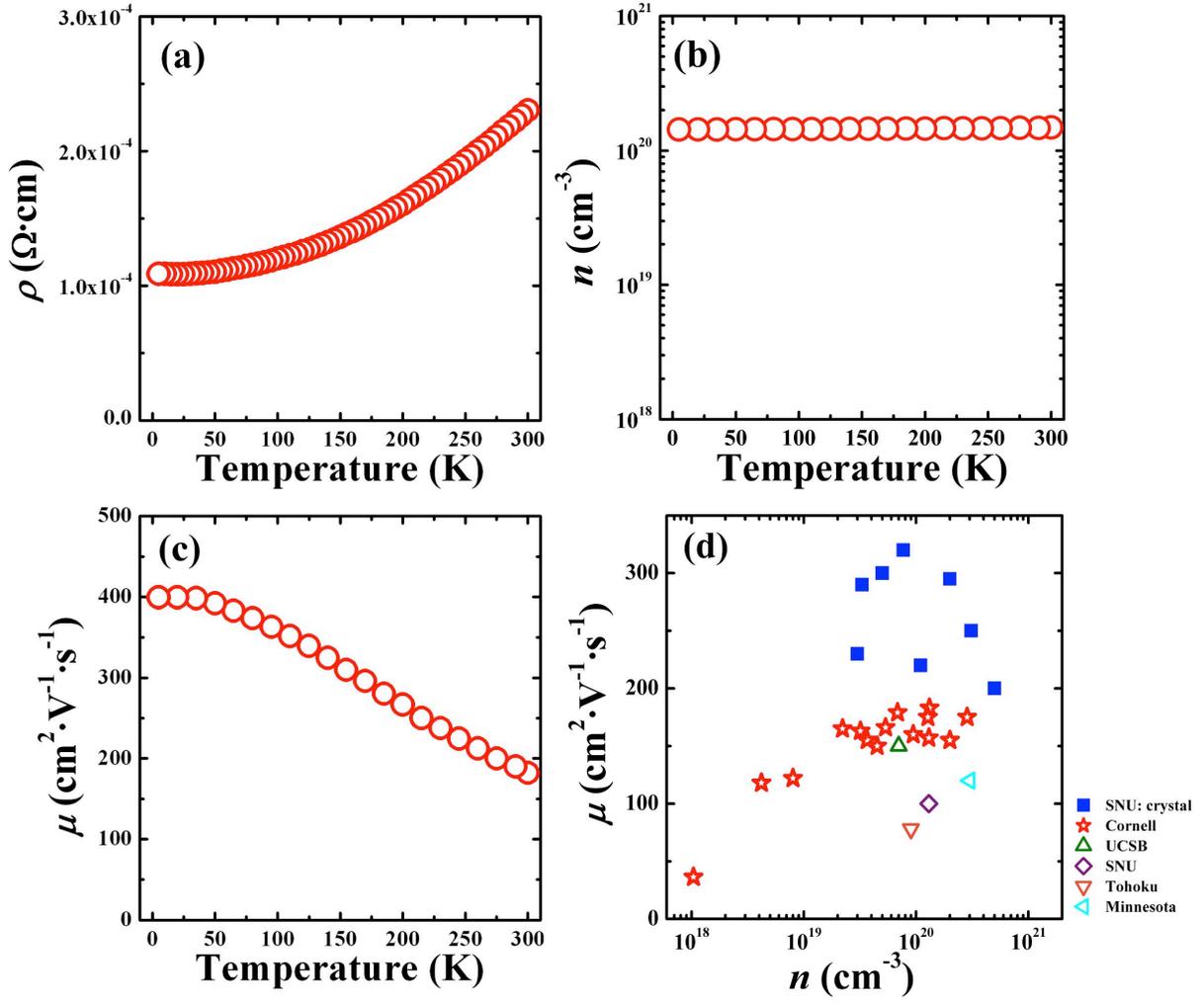

**Figure 5.**

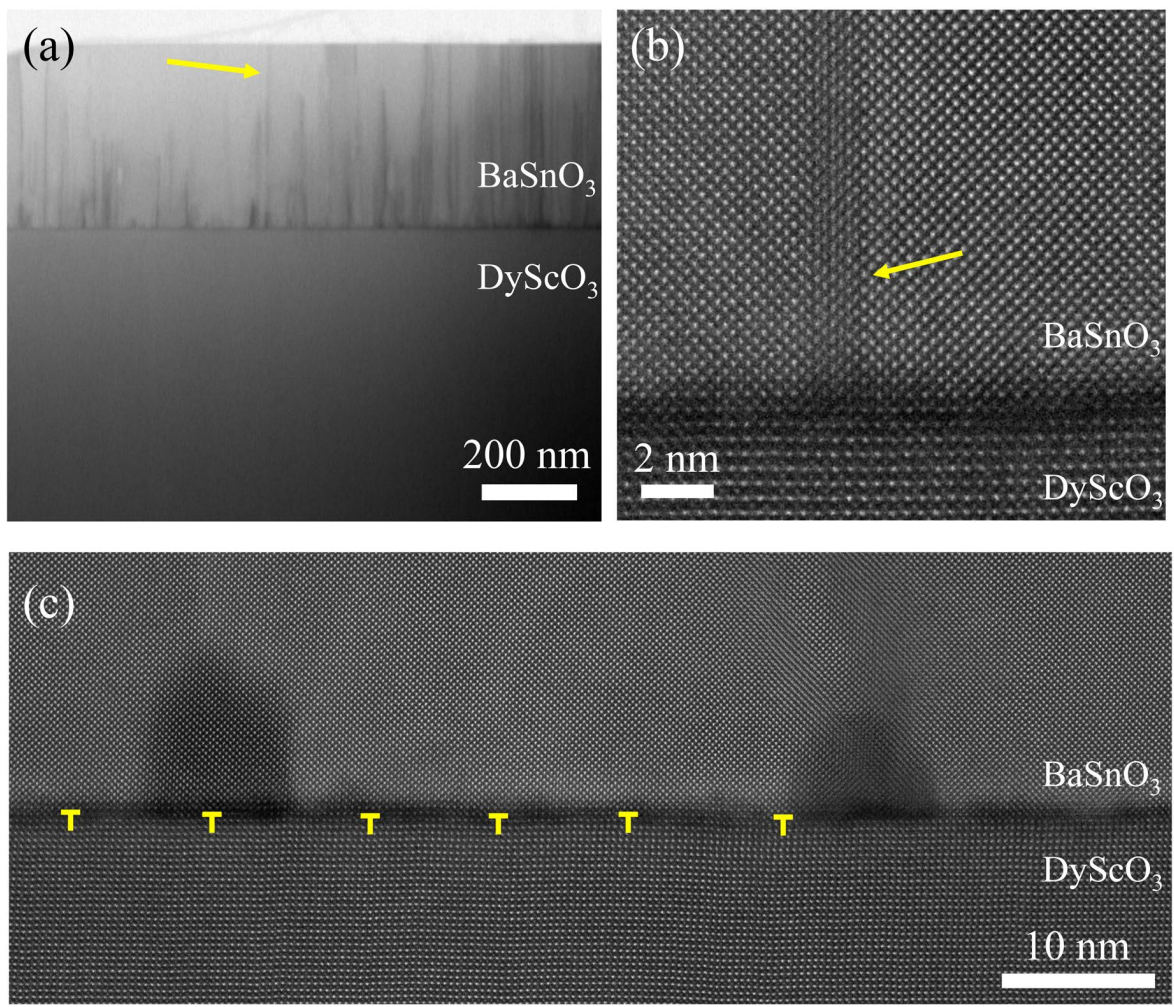

**Figure 6.**

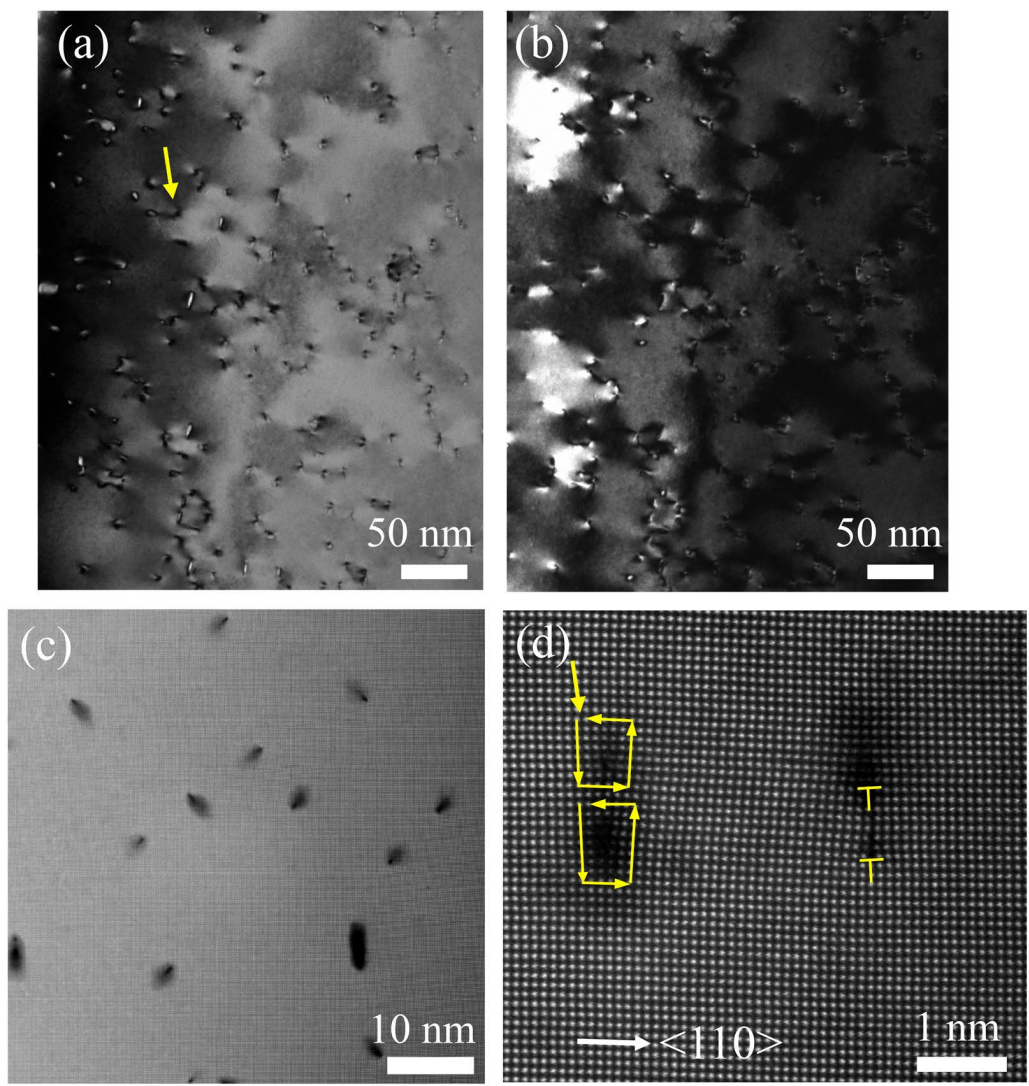